\documentclass{article}

\usepackage{PRIMEarxiv}

\usepackage[utf8]{inputenc} 
\usepackage[T1]{fontenc}    
\usepackage{hyperref}       
\usepackage{url}            
\usepackage{booktabs}       
\usepackage{amsfonts}       
\usepackage{nicefrac}       
\usepackage{microtype}      
\usepackage{lipsum}
\usepackage{fancyhdr}       
\usepackage{graphicx}       
\graphicspath{{media/}}     

\usepackage{url}
\usepackage{hyperref}

\usepackage{amsmath}
\usepackage{amssymb}

\usepackage{epsfig}
\usepackage{placeins}
\usepackage{subcaption}

\usepackage{url}

\title{Gaussian Blur and Relative Edge Response
}

\author{
  Austn C. Bergstrom, David Conran, David W. Messinger  \\
  Chester F. Carlson Center for Imaging Science\\
  Rochester Institute of Technology\\
  Rochester, NY\\
  \texttt{acb6595@rit.edu}
}

\begin{document}
\maketitle

\begin{abstract}
It is often convenient to use Gaussian blur in studying image quality or in data augmentation pipelines for training convoluional neural networks.  Because of their convenience, Guassians are sometimes used as first order approximations of optical point spread functions. Here, we derive and evaluate closed form relationships between Gaussian blur parameters and relative edge response, finding good agreement with measured results. Additionally, we evaluate the extent to which Gaussian approximations of optical point spread functions can be used to predict relative edge response, finding that Gaussian relationships provide a reasonable approximation in limited circumstances but not across a wide range of optical parameters.  
\end{abstract}

\keywords{Relative edge response \and Gaussian blur \and image quality}

\section{Introduction}

Relative edge response (RER) represents a convenient image quality metric summarizing the sharpness of an image by quantifying the spatial derivative of an image in the direction normal to an edge. Of note, RER is one of the three parameters used in the General Image Quality Equation used by the remote sensing community to quantify the utility of overhead images \cite{Harrington2015}.  

Because of the Gaussian distribution's mathematical simplicity, its ubiquity in image processing libraries \cite{Paszke2019, Virtanen2020, Tensorflow2015}, and its qualitative similarity to optical point spread functions, it is often convenient to use Gaussian kernels to blur images.  For instance, a range of studies on the relationship between image quality and deep learning have used Gaussian blur to manipulate image sharpness and spatial frequency content \cite{Dodge2016, Dodge2017, Geirhos2018, Hendrycks2019, Bergstrom2022}. 

Given the convenience and ubiquity of Gaussian blur kernels in studying image quality, we derive and evaluate closed form functions that approximate the relationship between Gaussian blur and RER. Additionally, we evaluate the extent to which we can approximate optical point spread functions with Gaussians for the purpose of quickly mapping system point spread functions to RER using the Gaussian RER relationships that we have derived. We observe that simple Gaussian approximations do not directly predict the RER that results from convolving an ideal edge with a realistic system point spread function. 

Specifically, we make the following contributions:
\begin{itemize}
    \item We derive simple, closed form relationships between RER and Gaussian blur.
    \item We verify our RER relationships using synthetic edge images blurred with Gaussian kernels.
    \item We show that Gaussian approximations of optical point spread functions yield images with RER that differs from the RER produced by the optical point spread functions themselves. 
\end{itemize}

\section{Derivations}\label{sec:rer-conv}

\subsection{First order approximation}

\begin{figure} [t!]
   \begin{center}
   \includegraphics[width=0.95\linewidth]{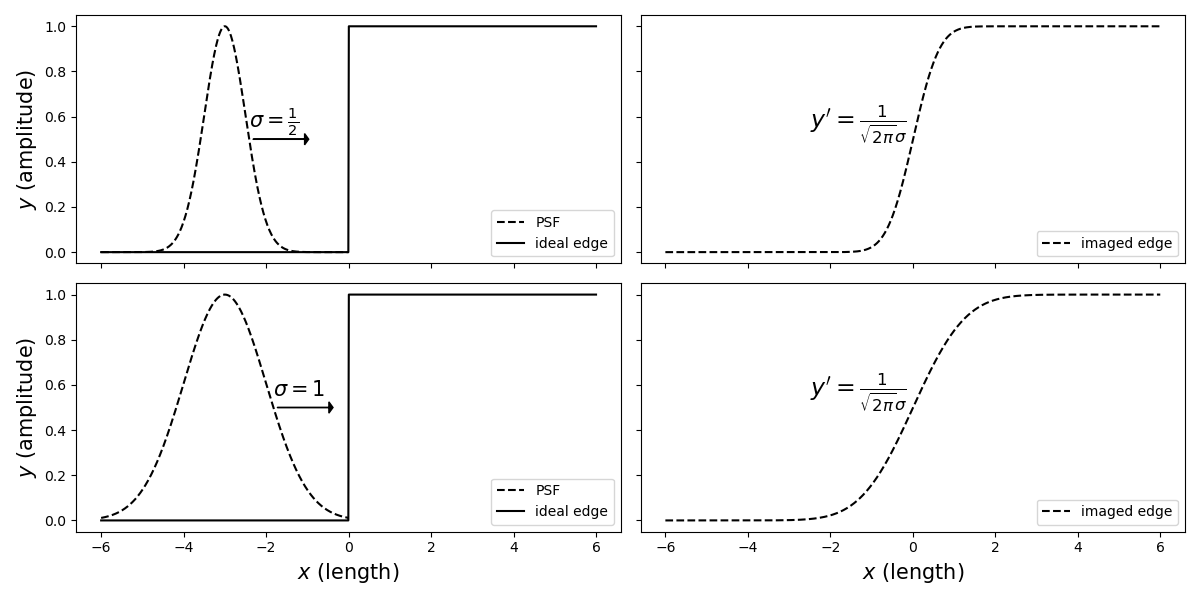}
   \end{center}
   \caption{Illustration of convolution and the effect of PSF width on relative edge response (RER).  To first order, RER is given by the slope of the image at the edge location. \label{fig:rer-conv}}
\end{figure} 

We begin by noting that we can approximate an imager as a linear shift invariant (LSI) system, allowing us to model the action of the system as a convolution of an input scene with the system's point spread function, where the convolution operation in given by 
\begin{equation}\label{eqn:convolution}
    g\left ( x \right )=f\left ( x \right )*h\left ( x \right )=\int_{-\infty }^{\infty}h\left ( \alpha \right )f\left ( x-\alpha \right )d\alpha=\int_{-\infty }^{\infty}h\left ( x - \alpha \right )f\left (\alpha \right )d\alpha.  
\end{equation}

To derive the relationship between Gaussian blur and RER, we consider a 1-dimensional image $g\left(x \right)$ formed by convolution of the edge object $f\left( x\right)$ with the the system point spread function $h\left(x\right)$. The edge object is described by the unit step function 
\begin{equation}\label{eqn:step_func}
    f\left( x\right) = \text{STEP}\left( x\right) = \begin{cases}
0 & \text{ where } x < 0 \\
\frac{1}{2} & \text{ where } x= 0\\
1 & \text{ where } x > 0 
\end{cases},
\end{equation}
and we approximate our point spread function  $h\left( x\right)$ as the normalized Gaussian
\begin{equation}\label{eqn:normed_gauss}
    h\left ( x \right )=\frac{1}{\sigma\sqrt{2\pi}}\exp\left ( \frac{-x^{2}}{2\sigma^{2}} \right ).
\end{equation}
Applying the definition of a convolution, we see that we can express our 1-dimensional image $g\left( x \right)$ with the integral
\begin{equation}\label{eqn:edge-image-integral}
    g\left( x\right)=\int_{-\infty}^{\infty}\frac{1}{\sigma\sqrt{2\pi}}\exp\left ( \frac{-\alpha^{2}}{2\sigma^{2}} \right )\text{STEP}\left ( x-\alpha \right )d\alpha.  
\end{equation}

Exploiting the properties of a step function and reversing the limits of integration for convenience, which is allowable since relative edge response is sign independent, we reach the expression for our edge image
\begin{equation}\label{eqn:edge-image-integral-simplified}
    g\left( x\right)=\int_{-\infty}^{x}\frac{1}{\sigma\sqrt{2\pi}}\exp\left ( \frac{-\alpha^{2}}{2\sigma^{2}} \right )d\alpha
\end{equation}

RER describes the sharpness of an image based on the slope of its edge response function; RER is a first order approximation of the spatial derivative of an image at the edge location \cite{Burns2000}.  In our 1D example, therefore, we have 
\begin{equation}\label{eqn:rer-gaussian-slope-only}
    \text{RER} \approx \frac{d}{dx}\int_{-\infty}^{x}\frac{1}{\sigma\sqrt{2\pi}}\exp\left ( \frac{-\alpha^{2}}{2\sigma^{2}} \right )d\alpha\bigg\rvert_{x=0}=\frac{1}{\sigma\sqrt{2\pi}}\exp\left ( \frac{-x^{2}}{2\sigma^{2}} \right )\bigg\rvert_{x=0}=\frac{1}{\sigma\sqrt{2\pi}},
\end{equation}
which is illustrated by Fig. \ref{fig:rer-conv}. 


\subsection{Refined approximation}\label{sec:updated-rer-model}

While RER is an approximation of the derivative of an image at the location of an edge, it is by definition a discrete approximation of this slope, measured by interpolating the edge spread function at $\pm \frac{1}{2}$ pixel \cite{Burns2000}.  Without accounting for the discrete sampling inherent in measurement, RER could approach infinity for sufficiently narrow point spread functions. To account for this sampling, we start wth edge image
\begin{equation}
    g\left( x\right)=\int_{-\infty}^{x}\frac{1}{\sigma\sqrt{2\pi}}\exp\left ( \frac{-\alpha^{2}}{2\sigma^{2}} \right )d\alpha
\end{equation}
formed by a system with a Gaussian PSF of standard deviation $\sigma$.  Applying the defintion of RER that accounts for discrete measurement, we have 
\begin{equation}
        \textup{RER} = g\left( 0.5\right) - g\left( -0.5\right),
\end{equation}
which is much less than $\frac{1}{\sigma\sqrt{2\pi}} $ as $\sigma \rightarrow 0$. To account for the effects of discrete sampling at $x=\pm 0.5$, we note that there is a convenient closed form relationship for $g\left ( 0.5 \right) - g\left ( 0.5 \right)$. If we re-write $g\left( x \right )$ as the sum of two integrals, with 
\begin{equation}
g\left( x\right)=\int_{-\infty}^{0}\frac{1}{\sigma\sqrt{2\pi}}\exp\left ( \frac{-\alpha^{2}}{2\sigma^{2}} \right )d\alpha + \int_{0}^{x}\frac{1}{\sigma\sqrt{2\pi}}\exp\left ( \frac{-\alpha^{2}}{2\sigma^{2}} \right )d\alpha, 
\end{equation}
we can discard first term since it is constant and falls out in subtraction, finding that 
\begin{equation}\label{eqn:RER-integral-subtraction}
    \textup{RER} = \int_{0}^{\frac{1}{2}}\frac{1}{\sigma\sqrt{2\pi}}\exp\left ( \frac{-\alpha^{2}}{2\sigma^{2}} \right )d\alpha - \int_{0}^{-\frac{1}{2}}\frac{1}{\sigma\sqrt{2\pi}}\exp\left ( \frac{-\alpha^{2}}{2\sigma^{2}} \right )d\alpha.
\end{equation}
In this form, we can see that RER is given by the difference of two scaled error functions, where the error function $\textup{erf}\left(x\right)$ is given by 
\begin{equation}\label{eqn:erf}
    \textup{erf}\left(x \right) = \frac{2}{\sqrt{\pi}}\int_{0}^{x}e^{-t^2}dt.  
\end{equation}
Using the change of variables 
\begin{equation}
    t = f \left( \alpha \right ) = \frac{1}{\sqrt{2}\sigma} \alpha, 
\end{equation}
we arrive at the expression 
\begin{equation}\label{eqn:RER-integral-change-of-vars}
    \textup{RER} = \frac{1}{\sigma \sqrt{2 \pi}} \left (\int_{0}^{\frac{1}{2\sqrt{2}\sigma}} e^{-\alpha^2} \sqrt{2}\sigma d\alpha - \int_{0}^{\frac{-1}{2\sqrt{2}\sigma}} e^{-\alpha^2} \sqrt{2}\sigma d\alpha \right), 
\end{equation}
which simplifies to 
\begin{equation}\label{eqn:rer-erf}
    \textup{RER} = \frac{1}{2} \left( \textup{erf} \left( \frac{1}{2\sqrt{2}\sigma} \right) - \textup{erf} \left( \frac{-1}{2\sqrt{2}\sigma} \right)\right) = \textup{erf} \left( \frac{1}{2\sqrt{2}\sigma} \right)
\end{equation}

\subsection{Extension to multiple blur stages}

In studying image quality, we are likely to be concerned with estimating the RER of images that have two distinct blur contributions, first by their system PSF and second in post-processing.   For a real image $g\left ( x \right)$, therefore, we have 
\begin{equation}\label{eqn:blurred-image-two-stage}
    g\left ( x \right )=f\left ( x \right )*h_{0}\left ( x \right )*h_{1}\left ( x \right )
\end{equation}
where $f$ is the object imaged, $h_0$ is the system PSF, and $h_1$ is the Gaussian blur kernel applied in post-processing. If we approximate the optical psf $h_0$ as a Gaussian of standard deviation $\sigma_0$, then 
\begin{equation}\label{eqn:sensor-psf-gaussian}
    h_0\left ( x \right )=\frac{1}{\sigma_{0}\sqrt{2\pi}}\exp\left ( \frac{-x^{2}}{2\sigma_{0}^{2}} \right ),
\end{equation}
\begin{equation}\label{eqn:secondary-psf-gaussian}
    h_1\left ( x \right )=\frac{1}{\sigma_{1}\sqrt{2\pi}}\exp\left ( \frac{-x^{2}}{2\sigma_{1}^{2}} \right ),
\end{equation}
and 
\begin{equation}\label{eqn:blurred-image-effective-psf}
    g\left ( x \right )=f\left ( x \right )*h_{effective}\left ( x \right ),
\end{equation}
where 
\begin{equation}\label{eqn:psf-effective}
    h_{effective}\left ( x \right ) = h_{0}\left ( x \right ) * h_{1}\left ( x \right ).
\end{equation} 
Here, we can apply the filter theorem (discussed in detail in \cite{Easton2010}) and note that the Fourier transform of our image $\mathfrak{F}\left\{ g\left ( x \right )\right\} = G \left ( \xi \right)$ is given by
\begin{equation}\label{eqn:filter_theorem}
    G\left ( \xi \right )=\mathfrak{F}\left\{ f\left ( x \right )\right\}\cdot \mathfrak{F}\left\{ h_{effective}\left ( x \right )\right\}=F\left ( \xi \right )\cdot H_{effective}\left ( \xi \right),
\end{equation}
where $H_{effective}$ represents the effective optical transfer function and is given by
\begin{equation}\label{eqn:transfer_function_effective}
    H_{effective}\left ( \xi \right) = H_0\left ( \xi \right)\cdot H_1\left ( \xi \right).
\end{equation}
Using the Fourier properties of a Gaussian, we have 
\begin{equation}\label{eqn:native_transfer_function}
    H_{0}\left ( \xi \right ) =  \exp\left( {2\pi^2\sigma_0^2\xi^2} \right), 
\end{equation}
\begin{equation}\label{eqn:secondary_transfer_function}
    H_{1}\left ( \xi \right ) =  \exp\left( {2\pi^2\sigma_1^2\xi^2} \right),
\end{equation}
and
\begin{equation}\label{eqn:effective-transfer-function-full}
    H_{effective}\left ( \xi \right ) =  \exp\left( 2\pi^2 \left (\sigma_0^2 + \sigma_1^2 \right)\xi^2 \right).
\end{equation}
Having found the effective transfer function, we can find the effective point spread function by taking the inverse Fourier transform of the effective transfer function according to 
\begin{equation}\label{eqn:psf-inverse-fourier-transform}
    h_{effective}\left ( x \right) = \mathfrak{F}^{-1} \left \{ H_0\left ( \xi \right)\cdot H_1\left ( \xi \right) \right\}.
\end{equation}
Here, 
\[
h_{effective}\left ( x \right ) =  \mathfrak{F^{-1}}\left\{ \exp\left( 2\pi^2 \left (\sigma_0^2 + \sigma_1^2 \right)\xi^2 \right) \right\}
\]
\begin{equation}\label{eqn:psf-inverse-fourier-transform-details}
    = \frac{1}{\sqrt{2\pi \left( \sigma_0^2 + \sigma_1^{2} \right)}}\exp\left( \frac{x^2}{2\left (\sigma_0^2 + \sigma_1^2 \right)}\right)
\end{equation}
\[
=\frac{1}{\sqrt{2\pi } \sigma_{effective}}\exp\left( \frac{x^2}{2\sigma_{effective}^2}\right),
\]
where $\sigma_{effective}=\sqrt{\sigma_0^2 + \sigma_1^{2}}$.

Accordingly, for two stages of Gaussian blur, we can model the effective point spread function as a Gaussian of standard deviation $\sigma_{effective}$, where
\begin{equation}\label{eqn:effective-std}
    \sigma_{effective} = \sqrt{\sigma_0^2 + \sigma_1^2}.
\end{equation}


\section{Method}

\begin{figure}[t]
\centering

\begin{subfigure}[c]{0.24\textwidth}
  \centering
  \includegraphics[width=1\linewidth]{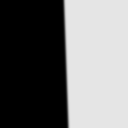}
  \caption{$\sigma=1$} 
\end{subfigure}
\begin{subfigure}[c]{0.24\textwidth}
  \centering
  \includegraphics[width=1\linewidth]{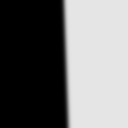}
  \caption{$\sigma=2$} 
\end{subfigure}
\begin{subfigure}[c]{0.24\textwidth}
  \centering
  \includegraphics[width=1\linewidth]{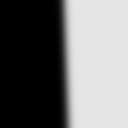}
  \caption{$\sigma=4$} 
\end{subfigure}
\begin{subfigure}[c]{0.24\textwidth}
  \centering
  \includegraphics[width=1\linewidth]{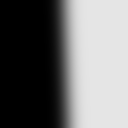}
  \caption{$\sigma=8$} 
\end{subfigure}

\medskip
\caption{Synthetic edge images with varied Gaussian blur.}\label{fig:edge_chips}
\end{figure}  

To evaluate the RER models derived above, we generated synthetic edge image chips (Fig. \ref{fig:edge_chips}) and measured RER using the slanted edge method outlined in \cite{Burns2000}, with our code available at \cite{bergstrom-rer-repo}. Specifically, we generated ideal slanted edges by defining the location of a near-vertical  in an $xy$-plane onto which to superposed our pixel grid.  We set pixels to the left of the edge equal to our dark value and pixels on the right side of the edge to our light value, and we assigned values to the border pixels according to the fraction of each on the light and dark side of the edge. Next, applied varying levels of Gaussian blur to our ideal edge image to generate edge images of varying RER.  Last, we down-sampled a subset of our edge images using integer pixel binning.  Table \ref{tab:multi-stage-no-downsample-chip-config} shows the blur parameters used for image chips without downsampling, and Tab. \ref{tab:multi-stage-downsampled-chip-config} displays the parameters used for image chips that were down-sampled after blurring. 

\begin{table}[h]
\begin{center}
    \captionof{table}{Parameters used in generating edge images with one and two-stage Gaussian blur and no down sampling}\label{tab:multi-stage-no-downsample-chip-config}
    \begin{tabular}{l r}
        \toprule
        First stage blur & 0.1 - 3 pixels \\
        Second stage blur & 0 - 3 pixels\\
        Combined blur \textit{(two stage images)} & 0.1 - 4.25 pixels\\
        \bottomrule
    \end{tabular}
\end{center}
\end{table}

\begin{table}[h]
\begin{center}
    \captionof{table}{Parameters used in generating edge images with two-stage Gaussian blur and integer down-sampling}\label{tab:multi-stage-downsampled-chip-config}
    \begin{tabular}{l r}
        \toprule
        First stage blur & 0.75 - 6 pixels \\
        Second stage blur & 0 - 6 pixels\\
        Combined blur \textit{(before down sampling)} & 0.75 - 8.5 (pixels)\\
        down sampling ratios & 2, 3, 4, 5 (dimensionless) \\
        Combined effective blur \textit{(after down sampling)} & 0.15 - 4.2 pixels\\
        \bottomrule
    \end{tabular}
\end{center}
\end{table}

We performed this down sampling in order to approximate the process of applying optical blur in the analog domain and then down-sampling with a focal plane array.  We used \textit{integer pixel binning} in order to avoid the effects of pixel interpolation.  For images with down-sampling applied, blur is linearly scaled by the down-sampling ratio. 

Finally, we assessed the extent to which Gaussian approximations of optical PSFs yielded equivalent RER values when used to blur synthetic edge chips. Do do so, we simulated system point spread functions using the code developed and described by Conran in \cite{conran-webinar}. Conran's model incorporates the optical system parameters shown in Tab. \ref{tab:psf-simulation-params}, with our simulations encompassing the ranges shown. 

\begin{table}[h]
\begin{center}
    \captionof{table}{System parameters used in optical PSF simulation}\label{tab:psf-simulation-params}
    \begin{tabular}{l r}
        \toprule
        f-number & 20 (dimensionless) \\
        aperture fill factor & 0.8 (dimensionless)\\
        pixel pitch & 8 $\mu m$ \\
        wavelength & 0.8 $\mu m$ \\
        wavefront error & 0.025 - 0.135 $\mu m$ \\
        smear & 0.05 - 0.15 pixel \\
        rms jitter & 2.6e-5 - 5e-4 pixel \\
        down sampling ratios & 1, 2 (dimensionless) \\
        \bottomrule
    \end{tabular}
\end{center}
\end{table}

For each optical PSF generated, we found the nearest two dimensional Gaussian using a non-linear least squares fitting routine. For each simulated optical PSF and its Gaussian best fit sibling, we blurred an ideal synthetic edge and measured resulting RER. 

\section{Results and Analysis}

\subsection{Gaussian Point Spread Functions}

\begin{figure}[t!]
\centering

\begin{subfigure}[c]{0.49\textwidth}
  \centering
  \includegraphics[width=1\linewidth]{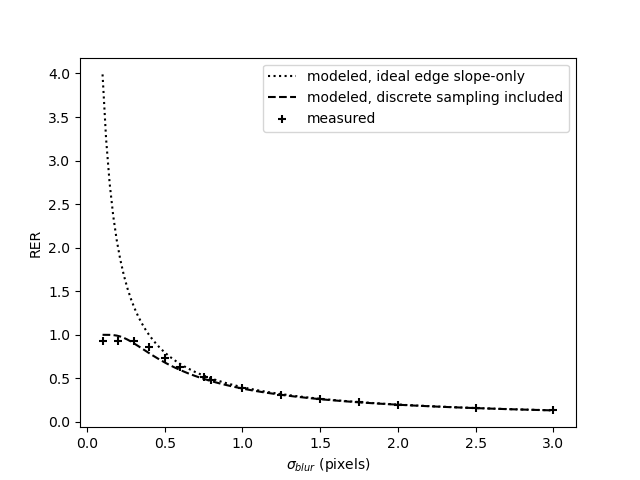}
  \caption{$0.1\leq\sigma\leq3$, single blur stage} 
\end{subfigure}
\begin{subfigure}[c]{0.49\textwidth}
  \centering
  \includegraphics[width=1\linewidth]{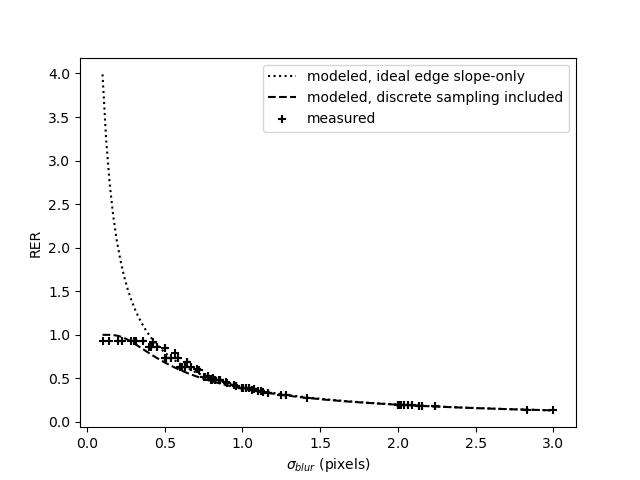}
  \caption{$0.1\leq\sigma_{effective}\leq3$, two blur stages} 
\end{subfigure}

\begin{subfigure}[c]{0.49\textwidth}
  \centering
  \includegraphics[width=1\linewidth]{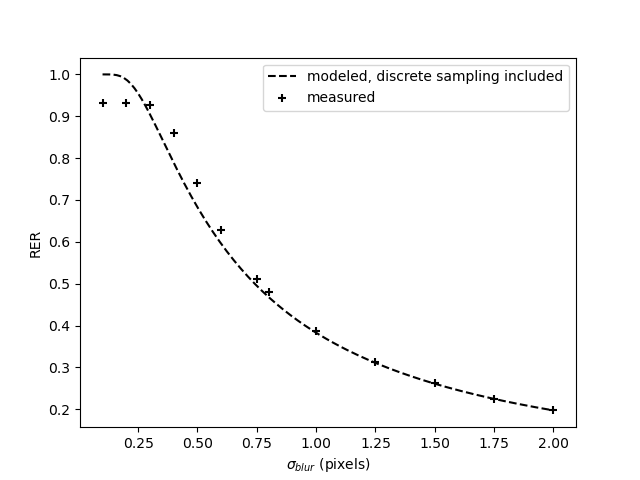}
  \caption{$0.1\leq\sigma\leq2$, single blur stage} 
\end{subfigure}
\begin{subfigure}[c]{0.49\textwidth}
  \centering
  \includegraphics[width=1\linewidth]{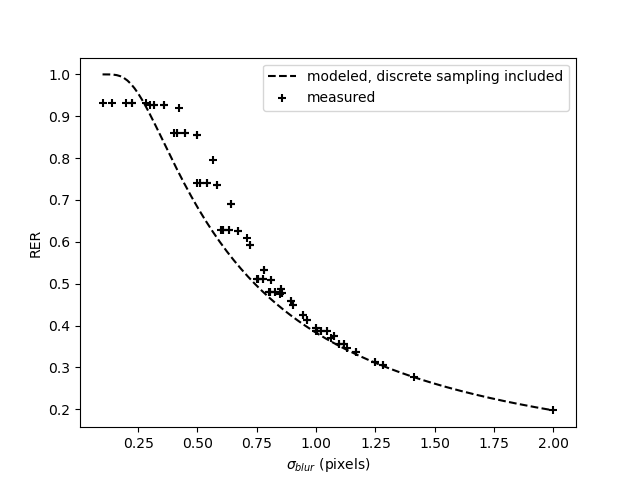}
  \caption{$0.1\leq\sigma_{effective}\leq2$, two blur stages} 
\end{subfigure}
\medskip
\caption{Relative edge response modeled and measured.}\label{fig:rer-no-downsampling}
\end{figure}

For synthetic edge images blurred with Gaussian kernels, we observed good agreement between our ideal edge slope model in Eqn. \ref{eqn:rer-gaussian-slope-only} for $\sigma \gtrsim 1$ pixel, with predicted RER exploding as $\sigma \rightarrow 0$. Our model incorporating discrete sampling in Eqn. \ref{eqn:rer-erf} avoids the small $\sigma$ catastrophy of the first model but still does not fit the data particularly well for $\sigma < 1$.  Figure \ref{fig:rer-no-downsampling} depicts these fits for modeled and measured RER using both of these models for edges blurred once and for edges blurred in two stages, where the combined standard deviation $\sigma_{effective}$ is calculated according to Eqn. \ref{eqn:effective-std} for the edges blurred twice. 

\begin{figure} [t!]
   \begin{center}
   \includegraphics[width=0.95\linewidth]{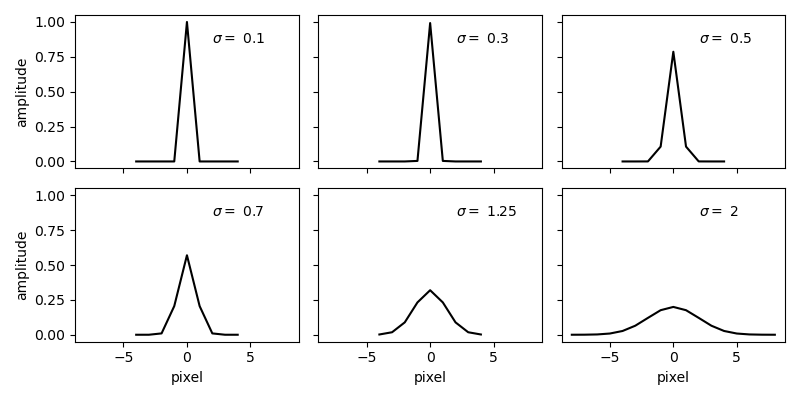}
   \end{center}
   \caption{Gaussian kernel profiles from the Torchvision library.  Here, we see that because of discrete sampling, blur kernels lose their Gaussian character for small $\sigma$ \label{fig:torchvision-gaussian-kernels}}
\end{figure}

Two factors explain the relatively poor performance performance at small $\sigma$ of our Eqn. \ref{eqn:rer-erf} model.  First, the model over-predicts RER for very small $\sigma$ because it neglects the  the transfer function of the pixels themselves.  Second, at very small $\sigma$, our blur kernels cease to be Gaussian in character due to the discrete sampling inherent in kernel generation.  Figure \ref{fig:torchvision-gaussian-kernels} shows the 1-dimensional profiles of the blur kernels from the Torchvision library that we used in generating our edge images.  As standard deviation $\sigma$ approaches 0, our blur kernels lose their Gaussian character and approach discrete delta functions. This non-Gaussian character of our blur kernels tends to drive RER up for small $\sigma$, leading our \ref{eqn:rer-erf} model to under-predict RER for moderately small $\sigma$. Because of this effect, our simplest model in Eqn. \ref{eqn:rer-gaussian-slope-only} yields the best prediction for RER when $\sigma > 0.5$ pixels and the images have not been down-sampled. 

\begin{figure}[t]
    \centering
    \includegraphics[width=0.6\linewidth]{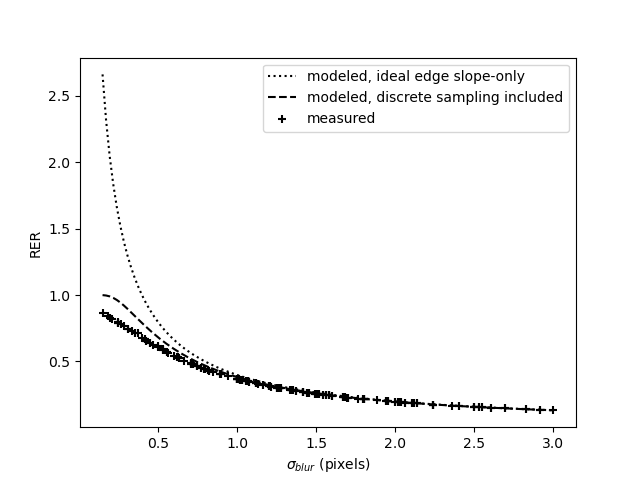}
    \caption[RER as a function of Gaussian blur, measured and predicted by ideal edge slope model and discrete sampling model]{RER as a function of Gaussian blur, measured and predicted by ideal edge slope model (Eqn. \ref{eqn:rer-gaussian-slope-only}) and discrete sampling model (Eqn. \ref{eqn:rer-erf}). These }
    \label{fig:rer-ideal-edge-slope-and-measured}
\end{figure}

To work around the effects of discrete blur kernels, we next consider synthetic edge images blurred at high resolution and then down-sampled. 
With down-sampling after blurring, our final Gaussian effective standard deviation is scaled by the same ratio as the image, enabling larger kernel standard deviations and therefore kernels of a more Gaussian character before down-sampling occurs.  For our synthetic edge images blurred this way to better approximate optical imaging, we observed that our simplest model (Eqn. \ref{eqn:rer-gaussian-slope-only}) still performs reasonably well for $\sigma > 1$ pixel, above which we avoid the problems inherent in the $1 / \sigma$ relationship.  We also see that our Eqn. \ref{eqn:rer-erf} discrete sampling now systematically over predicts RER at all small $\sigma$, as shown in Fig. \ref{fig:rer-ideal-edge-slope-and-measured}. 


We can understand this divergence between RER as measured and RER as predicted by the discrete sampling model at low $\sigma$ by recognizing that the transfer function of the pixels themselves significantly impacts RER at low blur values. Importantly, this transfer function is \emph{distinct} from the discrete sampling effects considered in \ref{sec:updated-rer-model} where we derived the discrete sampling model. While the discrete sampling model accounts for the impact of approximating the derivative of the edge spread function by measuring discretely at $\pm 1/2$ pixel, the pixel transfer function significantly changes the edge spread function by averaging the image signal across the width of the pixel.


To estimate the impact of the pixel transfer function, we calculated the combined transfer function 
\begin{equation}
    H_{combined}\left (\xi; \sigma \right) = H_{Gauss}\left (\xi ; \sigma\right) \cdot H_{pixel}\left (\xi \right),  
\end{equation}
for a range of $\sigma$ values.  Because we have specified $\sigma$ in units of pixels, we can conveniently treat our pixels as being unit width, yielding the pixel transfer function
\begin{equation}
    H_{pixel}\left (\xi \right) =  \mathfrak{F} \left \{ \textup{RECT} \left(x \right)\right \} = \textup{SINC} \left (\xi  \right), 
\end{equation}
where $\textup{RECT} \left(x \right)$ and $\textup{SINC} \left (\xi  \right)$ are a Fourier transform pair, expressed by 
\begin{equation}\label{eqn:rect}
\textup{RECT}\left(\xi \right) \equiv
\begin{cases}
        1 & \textup{ when } -\frac{1}{2} < \xi < \frac{1}{2} \\
        \frac{1}{2} & \textup{ when } \xi = \pm \frac{1}{2} \\
        0 & \textup{otherwise} \\
\end{cases}
\end{equation}
and 
\begin{equation}\label{eqn:sinc}
    \textup{sinc}\left(x \right) \equiv \frac{\textup{sin}\left(\pi x\right)}{\pi x}
\end{equation}
respectively. The transfer function of a Gaussian is a second Gaussian with standard deviation inversely proportional to the first Gaussian's standard deviation \cite{Easton2010}, leading to transfer function
\begin{equation}\label{eqn:gaussian-psf-transfer-function}
     H_{Gauss} = \mathfrak{F} \left\{ \frac{1}{\sigma\sqrt{2\pi}}\exp\left ( \frac{-x^{2}}{2\sigma^{2}} \right ) \right \} = \exp\left( {2\pi^2\sigma^2\xi^2} \right)
\end{equation}
for a Gaussian PSF of standard deviation $\sigma$. For a Gaussian PSF of standard deviation $\sigma$, therefore, we have a combined transfer function 
\begin{equation}\label{eqn:combined-transfer-function}
     H_{combined}\left (\xi; \sigma \right) = \exp\left( {2\pi^2\sigma^2\xi^2} \right) * \textup{SINC} \left (\xi  \right).
\end{equation}

\begin{figure}[ht!]
\centering

\begin{subfigure}[c]{0.49\textwidth}
  \centering
  \includegraphics[width=1\linewidth]{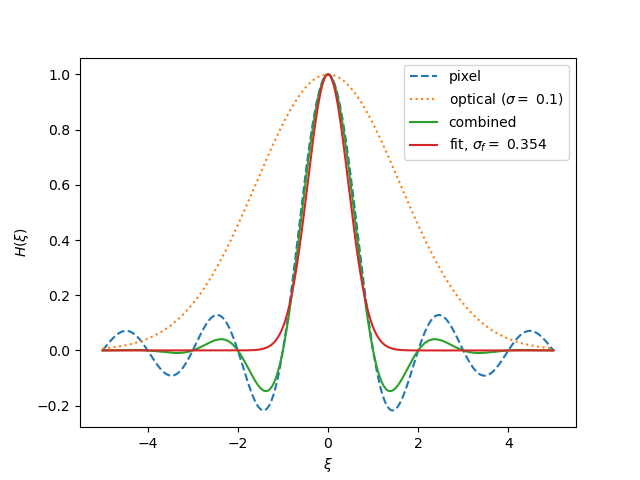}
  \caption{$\sigma=0.1$, $\sigma_{f}=0.354$} 
\end{subfigure}
\begin{subfigure}[c]{0.49\textwidth}
  \centering
  \includegraphics[width=1\linewidth]{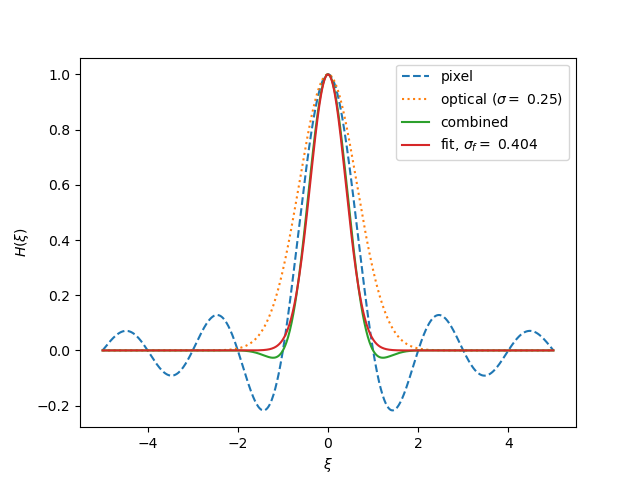}
  \caption{$\sigma=0.25$, $\sigma_{f}=0.404$} 
\end{subfigure}

\begin{subfigure}[c]{0.49\textwidth}
  \centering
  \includegraphics[width=1\linewidth]{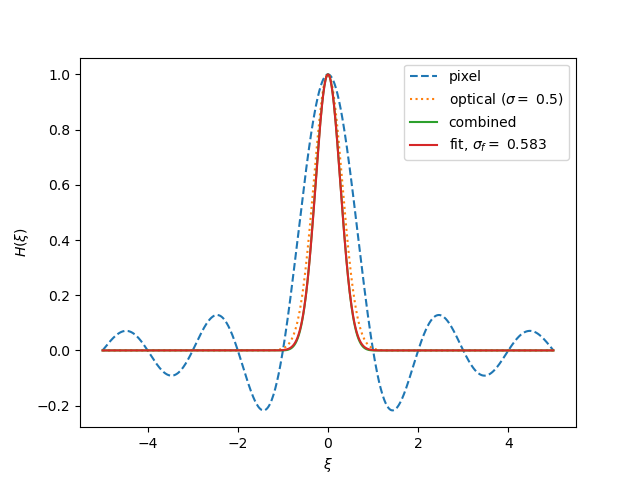}
  \caption{$\sigma=0.5$, $\sigma_{f}=0.583$} 
\end{subfigure}
\begin{subfigure}[c]{0.49\textwidth}
  \centering
  \includegraphics[width=1\linewidth]{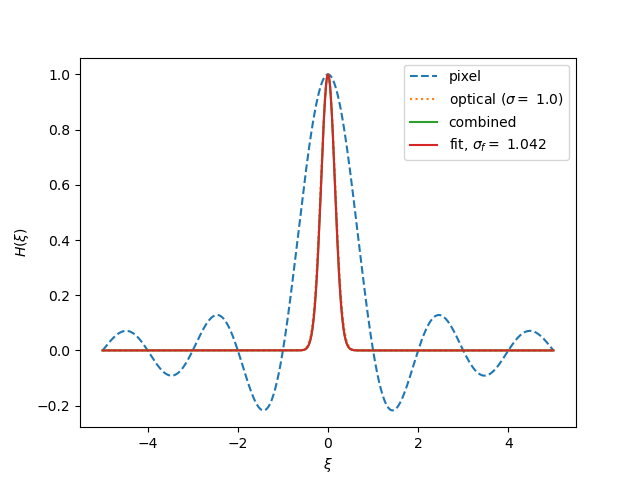}
  \caption{$\sigma=1$, $\sigma_{f}=1.042$} 
\end{subfigure}
\medskip
\caption[Combined transfer functions for varying Gaussian PSF widths and unit width pixels.]{Combined transfer functions for varying Gaussian PSF widths and unit width pixels. Note that in all cases $\sigma$ refers to the standard deviation of a normalized Gaussian PSF, whereas it is inversely proportional to the width of the Gaussian transfer function (see Eqn. \ref{eqn:gaussian-psf-transfer-function}). Here, we see that the pixel transfer function begins to significantly change the combined transfer function for $\sigma < 0.5$ pixels.}\label{fig:combined-transfer-functions}
\end{figure}  

\begin{figure}[t]
    \centering
    \includegraphics[width=0.5\linewidth]{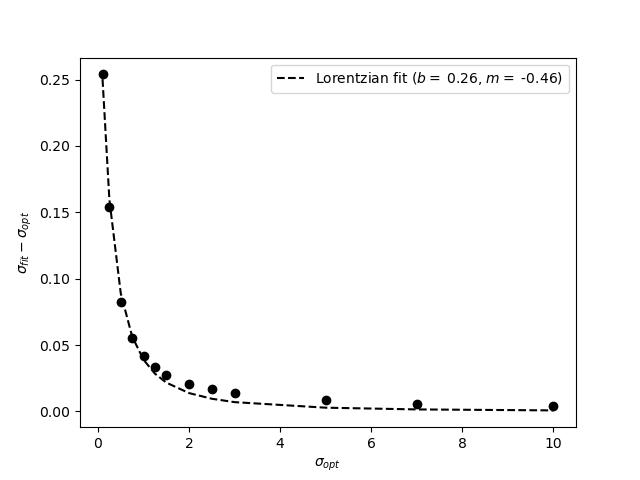}
    \caption[Residuals between original Gaussian $\sigma$ and the best fit $\sigma_f$ for the combined transfer functions]{Residuals between original Gaussian $\sigma$ and the best fit $\sigma_f$ for the combined transfer functions given by \ref{eqn:combined-transfer-function}}
    \label{fig:transfer-function-residual-fit}
\end{figure}

Figure \ref{fig:combined-transfer-functions} illustrates the interactions between the two terms in this transfer function. For wide Gaussian PSFs with large $\sigma$, we have narrow Gaussian transfer functions, in which case the pixel transfer function has minimal impact.  Conversely, for narrow Gaussian PSFs with small $\sigma$, we have wide Gaussian transfer functions, in which case the pixel transfer function has a significant impact.  We fit a Gaussian of the form $H_f \left(\xi \right) = \exp\left( {2\pi^2\sigma_{f}^2\xi^2} \right)$ to each combined transfer function $H_{combined}$ and observed that for small $\sigma$, the difference between the original PSF blur parameter and the best fit Gaussian blur parameter $\sigma_f$ varied significantly due to the impact of the pixel transfer function.  Figure \ref{fig:transfer-function-residual-fit} shows the difference between original $\sigma$ and $\sigma_f$ for varied $\sigma$, with the residuals following an approximately Lorentzian pattern, where the Lorentzian \cite{weisstein} is given by 
\begin{equation}\label{eqn:lorentzian}
    P\left(x \right) = \frac{1}{\pi}\frac{b}{\left(x - m \right)^2 + b ^2}.
\end{equation}

\begin{figure}[ht!]
\centering

\begin{subfigure}[c]{0.49\textwidth}
  \centering
  \includegraphics[width=1\linewidth]{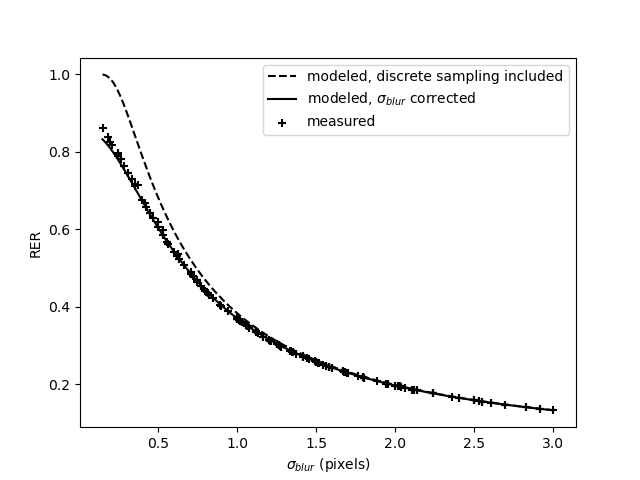}
  \caption{With and without blur correction} 
\end{subfigure}
\begin{subfigure}[c]{0.49\textwidth}
  \centering
  \includegraphics[width=1\linewidth]{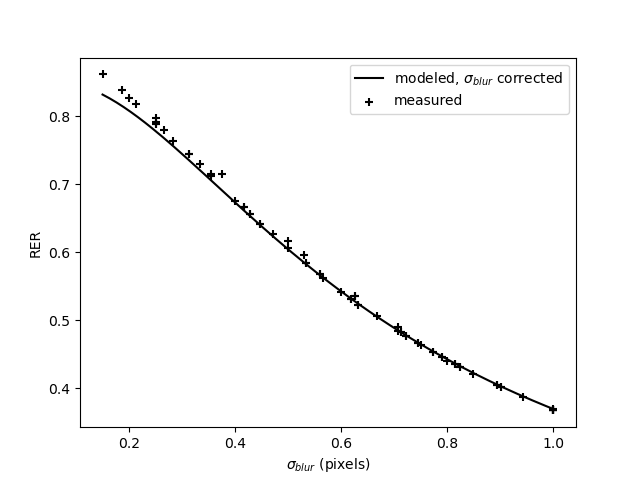}
  \caption{With blur correction, small $\sigma$} 
\end{subfigure}
\medskip
\caption[Combined transfer functions for varying Gaussian PSF widths and unit width pixels.]{RER as a function of Gaussian blur, measured and predicted by discrete sampling model (Eqn. \ref{eqn:rer-erf}) with and without pixel transfer function correction (Eqn. \ref{eqn:blur-correction-pix-xfer}) (left) The fully corrected model performs well down to roughly $\sigma \geq 0.25$. }\label{fig:rer-model-corrected-and-measured}
\end{figure}

Applying the fit parameters shown in Fig. \ref{fig:transfer-function-residual-fit} to Eqn. \ref{eqn:lorentzian}, we are able to estimate a combined Gaussian that accounts for the the Gaussian PSF as well as the pixel transfer function. Adding this correction to the original blur parameter $\sigma$, where
\begin{equation}\label{eqn:blur-correction-pix-xfer}
    \sigma_{corrected} = \sigma +  \frac{1}{\pi}\frac{b}{\left(\sigma - m \right)^2 + b ^2}, 
\end{equation}
and using the corrected blur value in the RER model given by Eqn. \ref{eqn:rer-erf}, we observe excellent agreement between modeled and measured RER across a wide range of blur parameter $\sigma$, as shown in Fig. \ref{fig:rer-model-corrected-and-measured}. 

Given the good agreement between modeled and measured RER, we can conclude that the model given by Eqn. \ref{eqn:rer-erf} can accurately predict RER as a function of Gaussian blur down to very low $\sigma$ if we apply a correction to account for the pixel transfer function. We highlight that the correction itself is found without reference to RER but is the result of finding the best fit Gaussian for the combined transfer function that results when a Gaussian PSF is convolved with a unit width pixel RECT function; our RER model is never fit to RER results, suggesting that the derivation from fist principles is sound. 

\FloatBarrier

\subsection{Gaussian Approximations of Optical Point Spread Functions}

\begin{figure}[h]
    \centering
    \includegraphics[width=0.6\linewidth]{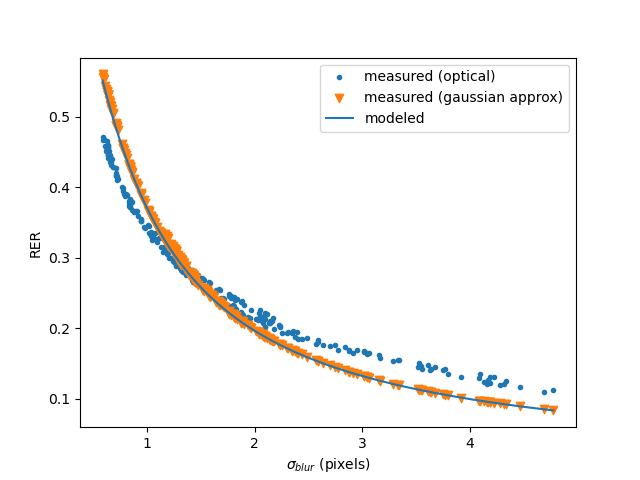}
    \caption[RER for images blurred with simulated optical PSFs and their best fit Gaussian approximations.]{RER for images blurred with simulated optical PSFs and their best fit Gaussian approximations. Simulated PSFs are from a combination of $Q=1$ and $Q=2$ systems with WFE, jitter, and smear values varying within the ranges shown in \ref{tab:psf-simulation-params}.}
    \label{fig:rer-opt-gauss-approx}
\end{figure}

Having established that we can predict the RER of a system with a purely Gaussian PSF, we next examined whether Gaussian approximations of simulated optical PSFs could be used to accurately predict RER.  Table \ref{tab:psf-simulation-params} shows the parameters used for our optical simulation.  We note that our simulated PSFs correspond to a $Q=2$ system before down-sampling, where $Q$ is the optical quality factor given by 
\begin{equation}
    Q = \frac{\lambda F}{p},
\end{equation}
for wavelength $\lambda$, f-number $F$, and pixel pitch $p$. \footnote{Conceptually, $Q$ is the ratio of the Airy radius to pixel pitch; it is also the ratio of detector cutoff frequency to optical cutoff frequency for a sensor with a 100\% fill factor.} After down sampling, our pixel pitch effectively doubles, which causes the drop in $Q$. Figure \ref{fig:rer-opt-gauss-approx} shows the RER that results from simulated optical PSFs and their best fit Gaussian approximations. 

From these results, we observe that the specific shape of optical kernels matters; the RER of an image blurred by one of our simulated optical kernels differs from the RER of an image blurred by its best-fit Gaussian kernel.  We note that we used other optical simulation parameters and observed similar differences (not shown here) between the RER of images blurred with our optical kernels and their least squares Gaussian siblings.

\section{Conclusion}

Here, we have derived closed form approximations for the relationship between Gaussian blur and relative edge response, and we have shown the limitations of these approximations. For images blurred at their final resolution, we find that our models do a reasonably good job of predicting RER down to blur standard deviations of approximately $0.5$ pixels, with our simplest relationship given by
\begin{equation}
    \text{RER} \approx \frac{1}{\sigma\sqrt{2\pi}}
\end{equation}
performing better than our refined approximation when $\sigma > 0.5$ but diverging rapidly as $\sigma \rightarrow 0$. 

For images blurred at high resolution and then down-sampled, which better approximates the process of optical blur and sampling by a focal plane, we find that our refined approximation, 
\begin{equation}
    \text{RER} \approx \textup{erf} \left( \frac{1}{2\sqrt{2}\sigma} \right), 
\end{equation}
yields good RER predictions for blur values greater than roughly 0.75 pixels, above which the Gaussian transfer function dominates the pixel transfer function. At smaller blur values, we can account for the pixel transfer function with the correction given by Eqn. \ref{eqn:blur-correction-pix-xfer}.  Additionally, when Gaussian blur is applied in two stages, we demonstrated that we can find the combined Gaussian standard deviation by combining the two standard deviations in quadrature, 
\begin{equation}
    \sigma_{combined} = \sqrt{\sigma_0^2 + \sigma_1^2}.
\end{equation}

Finally, we found that blurring with the least squares Gaussian approximation of an optical PSF does not yield the same RER as blurring with the optical PSF itself.


\bibliographystyle{spiebib.bst}
\bibliography{mendeley_bib.bib}

\end{document}